\documentstyle [aps,epsf,multicol]{revtex}
\renewcommand{\narrowtext}{\begin{multicols}{2}}
\renewcommand{\widetext}{\end{multicols}}
\begin{document}
\draft
\title{Photon Splitting in a Very Strong Magnetic Field }
\author{V. N. Baier, A. I. Milstein and R. Zh. Shaisultanov}
\address{Budker Institute of Nuclear Physics,
 630090 Novosibirsk, Russia}
\maketitle
\begin{abstract}
Photon splitting in a very strong magnetic field is analyzed
for energy $\omega < 2m$.
The amplitude obtained on the base of operator-diagram technique is used.
It is shown that in a magnetic field much higher than critical one the
splitting amplitude is independent on the field. Our calculation is in
a good agreement with previous results of Adler and in a strong contradiction
with recent paper of Mentzel et al.
\end{abstract}
\narrowtext

Virtual creation and annihilation of electron-positron pairs is
known to induce nonlinear self-action of an electromagnetic field.
Photon splitting in an external field is one of corresponding
processes of nonlinear QED.
Observation of a photon splitting is still a challenge for
experiment.

Theoretical study of this process has a rather long history.
Photon splitting in a constant and uniform external field was considered
in the beginning of 70th in [1-4], where earlier paper,
containing errors, were cited.
In \cite{1},\cite{3} the process was considered as a possible mechanism for
production of linearly polarized photons in a pulsar field
(assuming the field $H \sim H_0$).
At low photon energies ($\omega \ll m$,
$m$ is the electron mass, we set $\hbar=c=1$)
the splitting process can be analyzed
by using the Heisenberg-Euler (HE) effective Lagrangian.
In the weak field limit ($H \ll H_0$),
where $ H_0 = m^2/e = 4.41 \cdot 10^{13} G$ is
the critical magnetic field, the first term of expansion of
HE effective Lagrangian
can be used and the hexagon diagram contributes only.
This was done in
\cite{1},\cite{2}. The polarization selection rules, especially with allowance
for dispersion, were also obtained in \cite{1}, see also textbook \cite{5},
Sect. 129, 130 where the problem is given in detail.
The comprehensive investigation of the process under consideration was carried
 out
by Adler \cite{3}. For $\omega \ll m$ and an arbitrary field strength
the matrix element of the process was found
as a result of the application of the full HE effective Lagrangian.
At the same time, the allowed transition
amplitude was calculated for the general
case of an arbitrary field strength and photon
energy below pair creation threshold
($\omega < 2m$). A Green's function of the electron
in an external magnetic field in the Schwinger
proper-time representation was used.
Although the expression for this amplitude turned out to be
very unwieldy for application, the amplitude was calculated also numerically
in wide interval of magnetic fields $0 \leq H \leq H_0$
for $\omega=m$ as well as for $\omega \ll m$.
In \cite{4} photon splitting was considered in a crossed field
${\bf E} \perp {\bf H}, E=H$, using likewise the electron Green's function
in the proper-time representation.
Another form of the photon
splitting amplitude in a magnetic field
in a general case was obtained in \cite{6} using
similar approach for the Green's function calculation. Later,
photon splitting in a constant and uniform
electromagnetic field for arbitrary values of
both field invariants was considered in \cite{7}.
The operator diagram technique developed by
Katkov, Strakhovenko and one of us \cite{8} was used. As a result,
the solution of this technically quite cumbersome problem
was substantially simplified. The amplitudes obtained in particular
case of zero electric (or magnetic) field were found to be noticeably
compact than those obtained in \cite{3}. The results of
\cite{7} for $\omega \ll m$ and $H \ll H_0$
agree with those of \cite{1}-\cite{3}. In \cite{7} we performed numerical
calculations for the case $\omega \gg m$, because we were interested in
another possibility: photon splitting in electric fields of
single crystals at high energies \cite{7a}.
It is worth to note that in all mentioned paper
relativistic covariant and gauge invariant formulation of QED was used.

Recently, photon splitting was considered once more \cite{9}. That was
motivated by new astrophysics achievements. In this calculation
a non-covariant perturbation theory and Landau gauge were used. The results
of this paper as well as its subsequent application \cite{10} are in a strong
contradiction with all previous results.
Matrix element of photon splitting is found
in \cite{9} in the form of very cumbersome threefold infinite sum.
There is a series of short-comings in \cite{9}:
a) a low energy limit and weak field limit are not found from
general expression; b) the authors of \cite{9} suppose that
their approach is applicable for $H=H_0$, but
the result of calculation with the full HE
effective Lagrangian for $\omega \ll m$ is not reproduced;
c) photon dispersion in a magnetic field is not taken
into account; d) the study of cut off influence on summation
with respect to Landau level numbers is not sufficient because the
expression considered contains strong cancellations.
Adler \cite{11} criticized strongly
papers \cite{9} and \cite{10} and suggested to make the independent
calculation of photon splitting amplitude.

Because of the potential astrophysics implications of the process (see
e.g. \cite{12}) we
perform the numerical calculation and analysis of the expression for
photon splitting amplitude obtained using
another formulation of QED in external field \cite{7}.
We consider the most interesting region $\omega < 2m$ and arbitrary $H$.

Let photon with energy $\omega$ splits into two photons with energies
$\omega_1$ and $\omega_2$.
There is only one allowed transition in a magnetic
field \cite{1} with respect to photon polarizations:
$B \rightarrow CC$ in notations of \cite{7} or
$\perp \rightarrow \parallel \parallel$ in notations of \cite{1}.
Putting electric field $E=0$ in general representation for photon
splitting amplitude (eqs.(2.16)-(2.18),\cite{7}) we have for
allowed transition:

\widetext
\begin{equation}
\displaystyle{T=\frac{(4 \pi \alpha)^{3/2}\omega}{2 \pi^2} \int_{0}^{\infty}
 dx
\frac{\exp\left(-\frac{H_0}{H}x\right)}{x \sinh^2 x} \int_{0}^{x} dt_2
\Bigg[ \int_{0}^{t_2} dt_1 G\, \mbox{e}^{c \Phi} + \sinh^2t_2\,
\mbox{e}^{c \Phi_0}\Bigg]}
\label{1}\end{equation}
where
\begin{equation}
\begin{array}{rl}
&\displaystyle{c=\frac{H_0}{H}\left(\frac{\omega \sigma}{m} \right)^2,\quad
\Phi_0= \frac{t_2(x-t_2)}{x}-
\frac{\cosh x - \cosh(2t_2-x)}{2 \sinh x},~
\Phi= \big[
z_1z_2(t_1-t_2)(t_1-t_2+x)-z_1t_1(t_1-x)-}\\\\
&\displaystyle{z_2t_2(t_2-x)\big]/x -
\big[(1-z_1z_2)\cosh x-z_1\cosh(2t_1-x)-z_2\cosh(2t_2-x) +
z_1z_2\cosh (x+2t_1-2t_2)\big]/(2 \sinh x)},\\\\
&\displaystyle{G=\left[1-\left(z_1\cosh(2t_1-x)+z_2\cosh(2t_2-x) \right)
\cosh x \right]/x +
2cz_1z_2\sinh^2(t_2-t_1)\left[z_1 \sinh^2 t_1 +
z_2 \sinh^2 (t_2-x)\right]}.
\end{array}
\label{2}\end{equation}
\narrowtext
Here $\alpha=e^2=1/137$,~$z_{1,2}=\omega_{1,2}/\omega$,~
$\sigma = \sin \vartheta$,~
$\vartheta$ is the angle between direction of magnetic field and
momentum of the initial photon.
To derive (\ref{1}) we rotated the contour of
integration over each variable:~$x \rightarrow -ix$,
$t_{1,2} \rightarrow -it_{1,2}$.
This transformation is valid for any $\omega < 2m$. As a result,
the integrand in (\ref{1}) doesn't contain oscillating trigonometric
functions. So, it is convenient for numerical calculation.
It is easy to show that the amplitude $T$ is symmetric with respect to
interchange
of final photons ($\omega_1 \leftrightarrow \omega_2$).
By virtue of gauge invariance the amplitude
$T \propto \omega \omega_1\omega_2$ if
$\omega \rightarrow 0$ and also $T \rightarrow 0$ if $\omega_1 \rightarrow 0$
for any $\omega$.
That means that there are very strong compensations in (\ref{1}) and
this circumstance should be taken into account under numerical integration.
To overcome the difficulties it is convenient to
perform subtraction in the integrand of (\ref{1}):
$\mbox{e}^{c\Phi} \rightarrow \mbox{e}^{c\Phi}-1$
for the first term of $G$ (proportional to $1/x$)
and $\mbox{e}^{c\Phi_0} \rightarrow \mbox{e}^{c\Phi_0}-1$.
The sum of the subtracted terms is equal to zero.

For $\omega \ll m$ the main contribution to the amplitude is given by
the domain
of variables where $c\Phi \ll 1$, $c\Phi_0 \ll 1$. Then
expanding the corresponding exponents
and keeping linear in $c$ terms one can take the integrals over
$t_1$ and $t_2$. The result coincides with the photon splitting amplitude
found with the use of full HE effective Lagrangian (eq.(22) in \cite{3}).

For $\omega \sim m$ the integrals in (\ref{1}) are calculated numerically.
Without loss of generality one can put $\sigma=1$.
In Fig.1 the dependence of the amplitude $T$ on the final photon energy is
shown for $H=H_0/2$ (a) and $H=H_0$ (b) at different energies of the
initial photon.
The amplitude $T$ is normalized on $T_0$:
\[
T_0 = \frac{13}{315}\frac{(4\pi \alpha)^{3/2}}{\pi^2} \frac{\omega^3}{m^2}
\left(\frac{H}{H_0}\right)^3
\]
For $\omega/m = 0.1$ the result
coincides with
a very good accuracy (better than $10^{-3}$)
with the result
obtained from the full HE effective Lagrangian.

The total probability of allowed transition vs $w=\omega/m$ is shown in
Fig.2, where
\[ \displaystyle{W_0=T_0^2/(960 \pi \omega)}
=0.116 \left(\frac{\omega}{m}\right)^5
\left(\frac{H}{H_0}\right)^6~\mbox{cm}^{-1}.
\]
Curve (1) corresponds to $H/H_0=1$ and
curve (2) to $H/H_0=1/2$. Although the probability varies by
many orders of magnitude in the interval of parameters considered the
essential part of the variation is absorbed by $W_0$.
The function $W_0$ is nothing but the photon splitting probability
given by hexagon diagrams at $\omega \ll m$. Therefore, Fig.2 shows
the influence of higher order corrections with respect to
$\omega/m$ and $H/H_0$. The probability found with the use
of full HE effective Lagrangian also proportional to $(\omega/m)^5$. So,
the intersection points of the
curves with ordinate axis coincide with the probability $W_{HE}$
found in this approximation. One can see from Fig.2
that the probabilities $W$ and $W_{HE}$ are
essentially smaller than $W_0$ at $H \sim H_0$. At the same time,
$W/W_{HE}$ grows appreciably at $\omega \rightarrow 2m$. Therefore,
the exact photon energy-dependence should be taken into account.
Our numerical results agree (within a few percent)
with obtained that by Adler in paper \cite{3}
where the case $\omega=m$ was considered.

The behavior of the amplitude T in a very strong magnetic field $H \gg H_0$
is of evident interest from theoretical point of view.
In connection with this problem it is necessary to consider
the selection rules for photon splitting in a strong field.
Using the expression for eigenvalues of the photon polarization operator
in a magnetic field found in \cite{8}, eqs.(3.33)-(3.35)
we obtain that
$n_{\parallel}=1+\alpha/6 \pi $
and $n_{\perp}= \propto H$
for $H \gg H_0$
and $n_{\perp} > n_{\parallel}$ for any H.
This means (\cite{3,5}) that there is only
one allowed transition, which we considered above, for any field $H$.

For $H \gg H_0$ we found that the amplitude $T$
is independent of magnetic field and can be evaluated in analytic
form. The main contribution to two-fold integral in (\ref{1})
is given by the domain $x \sim H/H_0$ and $x-t_2 \sim 1$.
For threefold integral in (\ref{1}) the main contribution comes
from two domains: $x \sim H/H_0$,~$t_1 \sim H/H_0$, $x-t_2 \sim 1$;
and $x \sim H/H_0$,~$t_1 \sim 1$, $t_2 \sim H/H_0$. Performing
the corresponding expansion, we obtain
\widetext
\begin{equation}
\displaystyle{T(H \gg H_0)=T_1\frac{24m^4}{\omega^3}\bigg[
\frac{\omega_1} {\omega_2 \sqrt{4m^2-\omega_2^2}}
\mbox{arctan}\left(\frac{\omega_2}
{\sqrt{4m^2-\omega_2^2}}\right)
+ \frac{\omega_2}
{\omega_1 \sqrt{4m^2-\omega_1^2}}\mbox{arctan}
\left(\frac{\omega_1 }{\sqrt{4m^2-\omega_1^2}}\right)
-\frac{\omega}{4m^2}
\bigg]}
\label{3}\end{equation}
\narrowtext
where
\[
T_1=\frac{(4\pi \alpha)^{3/2}}{12\pi^2} \frac{\omega^3}{m^2}
\]
The amplitude calculated
using the full HE effective Lagrangian at $H \gg H_0$ is
\newline$T_{HE}=T_1 \omega_1 \omega_2/\omega^2$.
The dependence of the amplitude $T$ in this limit on the final photon
energy ($z_1=\omega_1/\omega$) is shown
in Fig.3 for different energies of the
initial photon. For strong
field and $\omega \rightarrow 2m$ one can see in Figs.1 and 3
a tendency of plato formation in the middle
of the distribution.

Thus, we performed the calculation  of photon splitting
amplitude using the exact formula valid
for any magnetic field $H$ and $\omega < 2m$.
If $\omega \ll m$ then our results coincide with the amplitude obtained
from the full HE effective Lagrangian. We obtained
that in a very strong field
$H \gg H_0$ the amplitude doesn't depend on a magnetic field.
We found that the refractive index $n_{\perp} > n_{\parallel}$ for any H.
So, there is only one allowed transition
$\perp \rightarrow \parallel \parallel$.
Therefore, a photon cascade could develop only if
magnetic field changes it's direction
(on distances much larger than the formation
length of photon splitting).
The results of our
calculation are in a good agreement with that obtained by Adler \cite{3} and
in a strong contradiction
with recent paper of Mentzel et al \cite{9}.


\end{multicols}
\newpage
\begin{figure}
\epsfxsize=3in
\epsffile{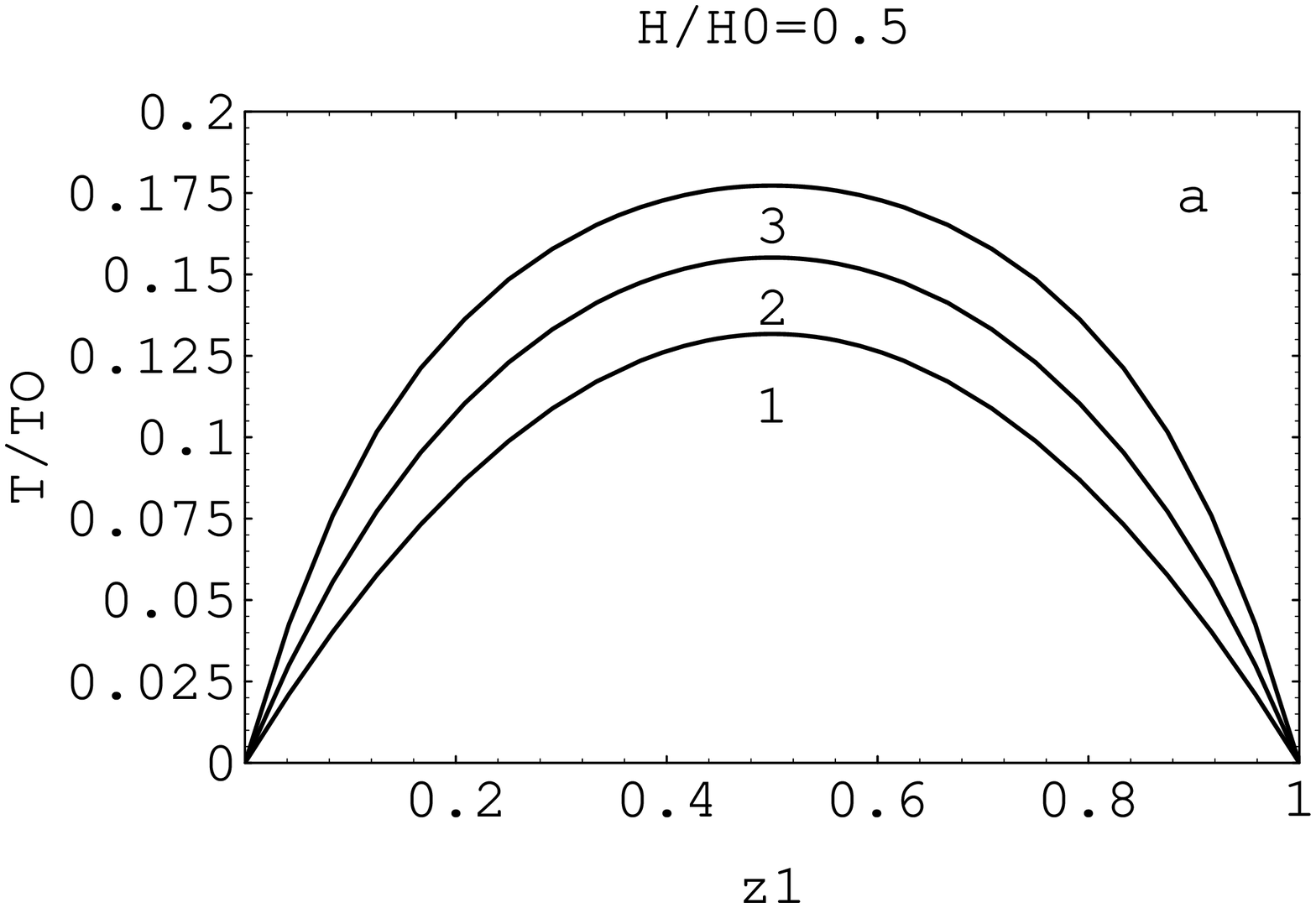}
\end{figure}
\begin{figure}
\epsfxsize=3in
\epsffile{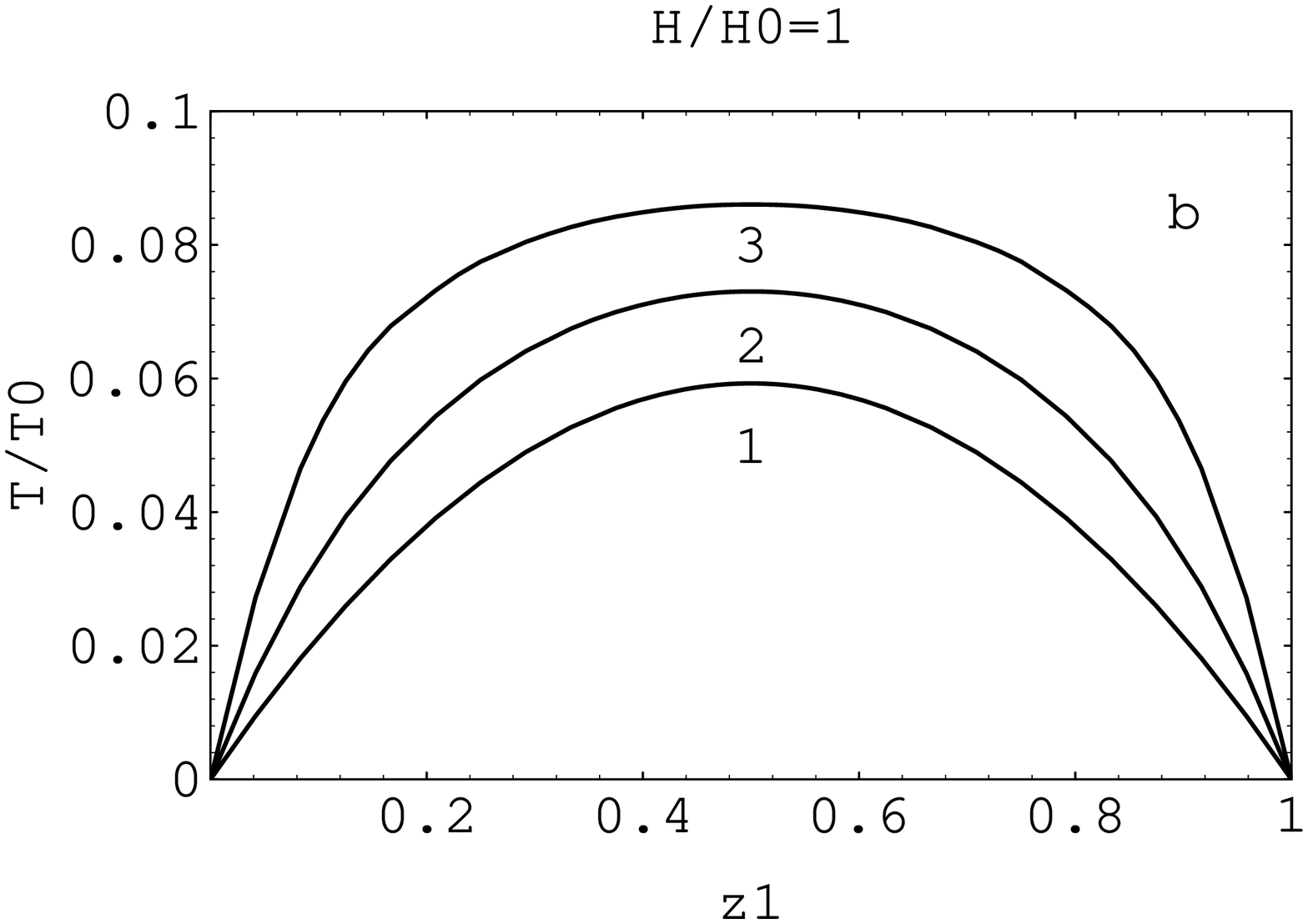}
\caption{\label{Fig.1} The dependence of photon splitting amplitude on the
final photon energy ($z_1=\omega_1/\omega$)
for $H=H_0/2$ (a) and $H=H_0$ (b) at different energies of the
initial photon: $\omega/m=0.1$(1),~$\omega/m=1.5$(2)~and
$\omega/m=1.9$(3). The amplitude $T$ is normalized on $T_0$ given in
the text.}
\end{figure}
\begin{figure}
\epsfxsize=3in
\epsffile{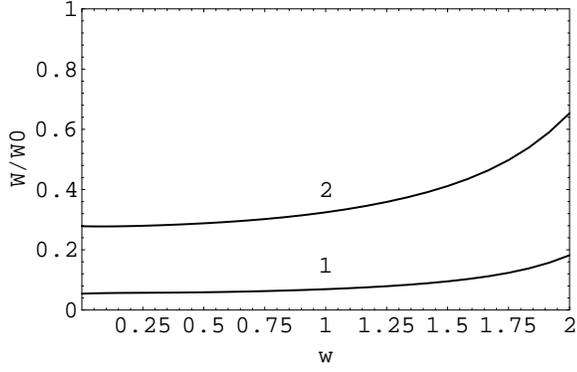}
\caption{\label{Fig.2} The dependence of total probability $W$
(in term of $W_0$) on photon energy ($w=\omega/m$) for $H=H_0$ (curve 1) and
$H=H_0/2$ (curve 2). The probability $W_0$ is given in the text.}
\end{figure}
\begin{figure}
\epsfxsize=3in
\epsffile{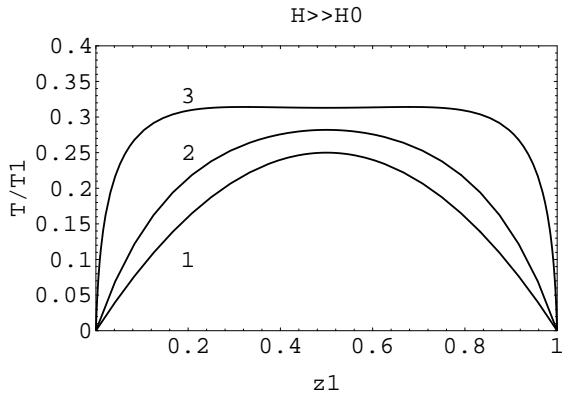}
\caption{\label{Fig.3} The dependence of the amplitude $T$ on the final photon
energy for $H \gg H_0$ for different energies of the initial photon:
$\omega/m=0.1$(1) ,~$\omega/m=1.5$(2)~and
$\omega/m=1.99$(3).}
\end{figure}

\end{document}